\documentclass[utf8]{clas} % for Health articles

\usepackage{hyperref,lineno,microtype,subcaption}
\usepackage[onehalfspacing]{setspace}
\usepackage{physics, soul}
\usepackage{siunitx}
\providecommand{\e}[1]{\ensuremath{\times 10^{#1}}}
\usepackage{graphicx}
\graphicspath{ {./images/} }
\def\keyFont{\fontsize{8}{11}\helveticabold }
\def\firstAuthorLast{Herrero Martin {et~al.}} %use et al only if is more than 1 author
\def\Authors{Clara Herrero Martin\,$^{1}$, Alon Oved\,$^{2}$, Rasheda A Chowdhury\,$^{3}$, Elisabeth Ullmann\,$^4$, Nicholas S Peters\,$^{3}$, Anil A Bharath\,$^{1}$ and Marta Varela\,$^{3,*}$}
\def\Address{$^{1}$Department of Bioengineering, Imperial College London, London, UK \\
$^{2}$Department of Computing, Imperial College London, London, UK \\
$^{3}$National Heart and Lung Institute, Imperial College London, London, UK \\
$^{4}$Department of Mathematics, Technical University of Munich, Munich, Germany }
\def\corrAuthor{Dr Marta Varela}

\def\corrEmail{marta.varela@imperial.ac.uk}

\begin{document}
\onecolumn
\firstpage{1}

\title[EP-PINNs]{EP-PINNs: Cardiac Electrophysiology Characterisation using Physics-Informed Neural Networks} 

\author[\firstAuthorLast ]{\Authors} %This field will be automatically populated
\address{} %This field will be automatically populated
\correspondance{} %This field will be automatically populated

\extraAuth{}

\maketitle
\begin{abstract}

\section{}
Accurately inferring underlying electrophysiological (EP) tissue properties from action potential recordings is expected to be clinically useful in the diagnosis and treatment of arrhythmias such as atrial fibrillation. It is, however, notoriously difficult to perform. We present EP-PINNs (Physics Informed Neural Networks), a novel tool for accurate action potential simulation and EP parameter estimation from sparse amounts of EP data. We demonstrate, using 1D and 2D \textit{in silico} data, how EP-PINNs are able to reconstruct the spatio-temporal evolution of action potentials, whilst predicting parameters related to action potential duration (APD), excitability and diffusion coefficients. EP-PINNs are additionally able to identify heterogeneities in EP properties, making them potentially useful for the detection of fibrosis and other localised pathology linked to arrhythmias. Finally, we show EP-PINNs effectiveness on biological \textit{in vitro} preparations, by characterising the effect of anti-arrhythmic drugs on APD using optical mapping data. EP-PINNs are a promising clinical tool for the characterisation and potential treatment guidance of arrhythmias.

\tiny
 \keyFont{ \section{Keywords:} cardiac electrophysiology, arrhythmia, machine learning, physics-informed neural network, atrial fibrillation, parameter estimation, action potential, electrogram analysis, optical mapping, systems identification, biophysical modeling, artificial intelligence} 
\end{abstract}

\section{Introduction}

Cardiac arrhythmias are extremely common pathologies caused by disturbances in the generation or propagation of electrical signals across the heart. 
Atrial fibrillation (AF), the most common sustained arrhythmia, affects 0.5\% of the world's population and accounts for 1\% of the NHS's total budget through its large impact on patient mortality and morbidity, especially stroke \cite{Hindricks2020}. Catheter ablation of atrial myocardium believed to host the sources of the arrhythmia is the mainstay of AF treatment, but its long-term efficacy is disappointing (54\%), especially in patients with persistent forms of the disease (43\%) \cite{Ganesan2013}.

The mechanisms behind AF are very complex, involving the interplay of several factors at different scales, from changes in membrane proteins to alterations in cardiac tissue composition and organ shape \cite{Nattel2017}. To characterise the arrhythmia, information about cardiac activity can be acquired by recording electrical potentials using electrodes placed on the chest (electrocardiogram, ECG) or, in a catheter lab, placed in direct contact with the myocardium (contact electrograms, EGMs). Expert analysis of these signals is extremely successful in the clinical diagnosis of arrhythmias and other types of cardiovascular disease \cite{Hindricks2020}. However, as sparse measurements of the combined electrical activity of large areas of the myocardium, electrical signals provide little direct information about local EP properties.

The ability to perform a detailed EP characterisation in the clinical setting could lead to improved treatments for arrhythmias. For example, evidence suggests that areas with abnormal EP properties (such as fibrotic or ischaemic regions) and their border-zone are often the sites of the abnormal electrical activity driving arrhythmias \cite{Nattel2017}. Cardiac regions characterised by EP changes such as low conduction velocity, heightened excitability or shortened action potential duration (APD) could be prime targets for localised therapies such as catheter ablation, likely improving their efficacy. So far, ablation strategies that target these EP heterogeneities have not been successful \cite{Calkins2012}, partly due to the difficulty in identifying suitable ablation sites.

In this study, we present EP-PINNs, a Physics-Informed Neural Network, as an artificial intelligence tool capable of inferring EP properties from sparse measurements of transmembrane potential, V, in cardiac tissue. We test EP-PINNs using \textit{in silico} data from EP biophysical simulations in several conditions and also \textit{in vitro} optical mapping data. EP-PINNs are deployed in forward mode, as high-resolution solvers of the biophysical equations that control EP systems, and also in inverse mode, as estimators of EP parameters. Tests are performed in 1D and 2D for single waves and spiral waves in homogeneous and heterogeneous conditions.
We further demonstrate a pharmacological application of EP-PINNs, as a tool to characterise the effect of two different channel blockers in \textit{in vitro} optical mapping data.

In the next section, we will introduce the EP biophysical model used and the PINNs technique, as applied to the EP problem. We will contextualise our work within the available techniques for parameter estimation in EP and other cardiovascular applications of PINNs.

\section{Background}
\subsection{Biophysical Models of Cardiac Electrophysiology}

Biophysical models of cardiac electrophysiology \cite{Clayton2011} are an important tool to understand how cardiac tissue properties affect the generation and propagation of cardiac electrical signals (action potentials, APs). They also offer an ideal means for the training and development of computational tools that may aim to infer EP properties from electrical and optical mapping measurements, such as EP-PINNs.

Several biophysical EP models have been proposed (see \href{https://models.cellml.org/electrophysiology}{models.cellml.org/electrophysiology}), each with varying degrees of detail aiming to reproduce different EP features, cardiac regions or animal/human experimental findings. Mathematically, these EP models usually take the form of a reaction-diffusion system where a diffusion term or equivalent \cite{Bueno-Orovio2014FractionalRepolarization.} models the propagation of the electrical signal across the cardiac tissue. In the monodomain formulation, a partial differential equation (PDE) describes the spatio-temporal variations in the electrical potential across a myocyte cell membrane ($V$). This PDE is usually coupled to one or more ordinary differential equations (ODEs) describing how, at each point in time and space, $V$ and other local state variables both determine and are determined by the flux of ions across the cell membrane \cite{Clayton2011}.

The most parsimonious model of the action potential describes it as a travelling excitation wave followed by a non-excitable (refractory) region. This representation requires at least two state variables: $V$, which spreads (diffuses) across neighbouring regions, and a non-observable, non-diffusible recovery variable $W$ which effectively controls the refractoriness and restitution properties of the model. One of the simplest models that captures these properties is the 6-parameter canine ventricular Aliev-Panfilov model \cite{Aliev1996}, which models the transmembrane ionic currents ($V - W$ relationship) using smooth, differentiable functions. Furthermore, the diffusion of $V$ across the cardiac tissue can be described by the monodomain equation \cite{Clayton2011}, which, when combined with the Aliev-Panfilov model gives:
\begin{align}
    \pdv{V}{t} &= \vec \nabla . (D \vec \nabla V) - k V (V-a) (V-1) - V W \label{eq:AlievP_V}\\
    \dv{W}{t} &= (\epsilon + \frac{\mu_1 W}{V + \mu_2}) (-W - k V (V-b-1)) \label{eq:AlievP_W}
\end{align}

The diffusion term $\vec \nabla . (D \vec \nabla V)$ reduces to $D \nabla^2 V$ in the case of homogeneous and isotropic conduction, i.e. when the diffusion tensor $D$ is approximated by the same scalar throughout. Intuitively, this term quantifies how fast $V$ is able to spread to its immediate neighbourhood to become more spatially homogeneous. $D$ is determined mostly by the electrical conductivity of the myocardium and is a strong determinant of the propagation velocity of the AP. To prevent a leakage of $V$ to regions outside the heart domain, the system described by Eq \ref{eq:AlievP_V}, \ref{eq:AlievP_W} usually obeys no flux Neumann boundary conditions: $\frac{\partial V}{\partial \vec{n}} = 0$ in the boundary of the heart tissue.

The $- k V (V-a) (V-1) - V W$ term in Eq \ref{eq:AlievP_V}, \ref{eq:AlievP_W} models the rapid changes in $V$ caused by ionic fluxes across the cell membrane. $a$ is related to the excitation threshold (i.e. the minimum $V$ value that leads to the onset of an AP). The model's APD and refractoriness can, in turn, be controlled using $b$. The values for each of the model parameters are typically chosen empirically to reproduce observed electrical signals - we use the values listed in Supplementary Table 1. The Aliev-Panfilov model uses rescaled units: $V$ is adimensional (typically in the $[0,1]~AU$ interval) and time is measured in temporal units, referred to as $TU$ throughout this study. $1~TU$ corresponds to approximately $13~ms$ \cite{Aliev1996}.

\begin{figure}[h]
\centering
\includegraphics[width=\columnwidth]{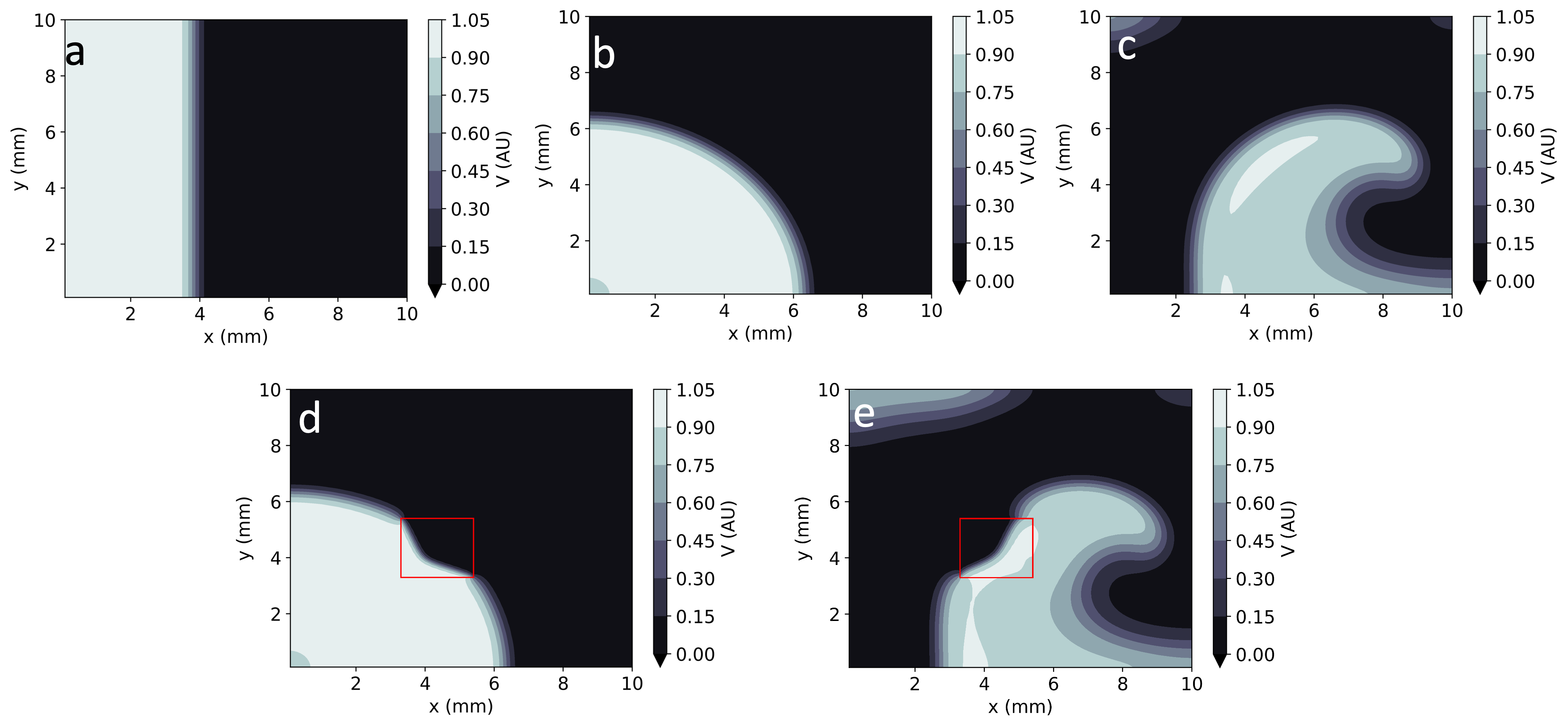}
\caption{Numerical solutions to the Aliev-Panfilov monodomain system for: \textbf{a)} Planar wave; \textbf{b)} Centrifugal wave; \textbf{c)} Spiral wave; \textbf{d)} Centrifugal wave in the presence of a square heterogeneity in $D$; \textbf{e)} Spiral wave in the presence of a square heterogeneity in $D$.}
\label{fig:Fig1}
\end{figure}

Other more complex EP models exist, through which it is possible to model individual membrane ionic currents and other biological components relevant for the AP and its propagation. One example used in the current study is a 14-current 30-variable canine atrial model that incorporates different degrees of EP remodelling caused by atrial fibrillation \cite{Varela2016}.

By assigning different sets of parameters and/or initial conditions to EP mathematical models, they can represent the electrical behaviour of the heart in both healthy and arrhythmic conditions. Healthy conditions are usually represented as unidirectional smooth propagation from a single source (Figure \ref{fig:Fig1}a,b). Large sources produce wave fronts which are close to planar, whereas point-like wavefronts lead to centrifugal (convex) wavefronts. Arrhythmias are usually modelled as one or more re-entrant waves (called spiral waves or rotors in 2D - Figure \ref{fig:Fig1}c). Moreover, the wavefronts and/or wavebacks of spiral waves can, in some instances, fragment (break-up), leading to complex activation patterns \cite{Fenton1998}. These models can also consider localised pathology such as fibrosis, scar or ischaemia as heterogeneities in one or more model parameters. Localised reductions in $D$, for instance, can be used to reproduce the slow-down of AP propagation in fibrotic lesions \cite{Roy2018}. These heterogeneities can lead to local changes in the curvature of the activation wavefront (Figure \ref{fig:Fig1}d,e).

\subsection{Parameter Estimation in Cardiac Electrophysiology}
Biophysical models are usually employed in forward mode, with the aim of reproducing the system's behaviour assuming that all model parameters are perfectly known. In many circumstances, such as the identification of pathology from EP measurements, it is more desirable to use these biophysical models in inverse mode, inferring the tissue parameters that underlie an observed system behaviour. This task, often called parameter estimation (or systems identification), is extremely challenging for several reasons.

The observed data are typically insufficient to identify a unique parameter value, since the observations are usually sparse, incomplete and polluted by noise. This can be handled by optimization methods, such as least squares, that fit data and model in an optimal way, combined with problem-dependent regularization terms to stabilize the estimate \cite{alma991000411625601591}. Typically, the optimization requires many forward runs of the underlying model. Unfortunately, in physical systems that are described by PDEs such as in EP, the forward runs are very computationally expensive in realistic 2D and 3D settings. These difficulties are exacerbated by the fact that the parameters in many EP models are heterogeneous. Hence, we often do not infer a scalar quantity but a (discrete) function in space and/or time. This increases the cost and complexity of the inverse mode even further.

Modern inverse problem solvers such as ensemble Kalman filters \cite{Hoffman2020} sequential Monte Carlo \cite{Drovandi2016} and parameter estimation based on Markov chain Monte Carlo \cite{Dhamala2018} are often combined with reduced order models \cite{Fresca2020DeepElectrophysiology} or multi-fidelity approaches \cite{SahliCostabal2019Multi-fidelityModels} to decrease the computational complexity of parameter estimation. These very complex methods typically make strong assumptions about the statistical distribution of model parameters and require dedicated problem-specific parameterisation. It is not clear how well they can generalise when applied to a different EP model or task.

\subsection{Physics Informed Neural Networks (PINNs)} \label{Intro_PINNs}
Physics Informed Neural Networks (PINNs) \cite{Raissi2019} are an exciting new tool for the study of physical systems modelled by PDEs and/or ODEs. PINNs have been shown to both efficiently find high-resolution solutions (forward problem) and perform parameter estimation (inverse problem) in a variety of systems \cite{Karniadakis2021}. As opposed to most types of neural networks (NNs), whose inputs are exclusively empirical data, PINNs incorporate explicit knowledge about the physical laws that govern a system. This allows PINNs to compute solutions to initial and/or boundary value problems with comparatively less training data than conventional NNs.

Very briefly, PINNs make use of automatic differentiation \cite{Baydin2018AutomaticSurvey} to rewrite the differential equations that a system obeys as the minimisation of a functional $f$. For example, for Eqs \ref{eq:AlievP_V}, \ref{eq:AlievP_W}, this functional would be defined as: 
\begin{align}
    f_V &= - \pdv{V}{t} + \vec \nabla . (D \vec \nabla V) - k V (V-a) (V-1) - V W\\
    f_W &= - \dv{W}{t} + (\epsilon + \frac{\mu_1 W}{V + \mu_2}) (-W - k V (V-b-1))
\end{align}

PINNs are trained to minimise a hybrid loss function, $L$, which ensures the system obeys known physical laws (described by $f_V$ and $f_W$), whilst simultaneously fitting known empirical system measurements. $L$ thus includes terms to account for:
\begin{itemize}
    \item agreement with the experimental measurements, $L_{data}$;
    \item consistency with the physical laws of the system, $L_{f_V} + L_{f_W}$;
    \item consistency with boundary $L_{V_{BC}}$ and initial value conditions $L_{V_{IC}}$.
\end{itemize}
Mathematically:
\begin{align}
    % \begin{multline}
    L &= L_{data} + L_{f_V} + L_{f_W} + L_{V_{BC}} + L_{V_{IC}} 
    \label{eq:LossTerms}
    \\
    L &= \frac{1}{N} \sum_{i=1}^{N} (V({x_i},t_i)- {V_{GT}}_i)^2 + \frac{1}{N_f} \sum_{j=1}^{N_f} (f_V({x_j},t_j)^2 + f_W({x_j},t_j)^2) + \\
    & \frac{1}{N_b} \sum_{k=1}^{N_b} (\frac{\partial V}{\partial \vec{n}}(x_k,t_k))^2 + \frac{1}{N_0} \sum_{l=1}^{N_0} (V({x_l},t_0) - V_0) ^2
    \label{eq:Loss}
    % \end{multline}
\end{align}

Each of the terms of the loss function is typically computed in different domains:
\begin{itemize}
  \item $({x_i},t_i)$ are the $N$ measurement points, where ground truth (GT) experimental measurements, $V_{GT}$, are known and should be reproduced as closely as possible.
  \item $({x_j},t_j)$ are the $N_f$ residual points, where the fulfilment of the biophysical equations is tested.
  \item $({x_k},t_k)$ are the $N_b$ boundary points, where the network aims to fulfill the Neumann boundary condition for $V$.
  \item $({x_i},t_0)$ are the $N_0$ initial points, where the known initial condition, $V_0$, is replicated as closely as possible.
\end{itemize}

There is an asymmetry in $L$ between measurable ($V$) and latent ($W$) variables: only experimental measurements (and initial conditions) for $V$ are usually available. Moreover, as $W$ does not diffuse across the tissue, it does not obey any boundary conditions. 

PINNs have typically been used for two main purposes \cite{Karniadakis2021}. In the so-called forward mode, the NN's parameters are optimised to provide a representation of the physical system of interest consistent with observations. Using this representation, the system's differential equations can subsequently be solved at an arbitrarily high spatial or temporal resolution, bypassing the constraints (e.g. small temporal and spatial steps) of traditional numerical solvers. In inverse mode, PINNs additionally perform parameter inference (systems identification) by having the NN optimise one or more of the equation parameters (which here represent tissue EP properties) during the training process.

\subsection{Cardiovascular Applications of PINNs}
PINNs have recently been used in several areas of cardiovascular medicine, especially for applications related to blood flow. Examples include the estimation of myocardial perfusion and related physiological parameters from dynamic contrast enhanced MRI \cite{VanHerten2020} and the estimation of haemodynamic parameters from microscopic images of aneurysms-on-a-chip \cite{Cai2021}.

In the field of cardiac EP, Sahli-Costabal \textit{et al} used PINNs in forward mode to estimate activation time maps (ATs, i.e. the arrival times of the action potential) and conduction velocity (CV) maps in the left atrium at high spatial resolution \cite{SahliCostabal2020}. Sahli-Costabal's method uses PINNs to solve the (isotropic diffusion) eikonal equation, a simple relationship between ATs and the spatial gradient of CV. This effectively interpolates AT and CV across the left atrial surface. Although their PINNs implementation was exclusively deployed on simulated data, the proposed application is very clinically relevant, as it aims to mitigate the low spatial resolution of clinical AT measurements. Grandits \textit{et al} subsequently extended this PINNs-eikonal equation framework to anisotropic conduction, using it to estimate high-resolution AT maps and fibre directions from \textit{in silico} and patient data \cite{Grandits2021}. The PINNs method performance was nevertheless lower than that of a traditional (variational) inverse solver \cite{Grandits2020}. 
As they rely on the eikonal equation, these tools are not well suited to the study of arrhythmic conditions or to the inference of EP parameters other than AT and CV.

\subsection{Optical Mapping for Experimental EP-PINNs Testing}
Maps of transmural electrical potential ($V$), similar to those simulated using the Aliev-Panfilov model, can be experimentally recorded using optical mapping. Optical mapping is a technique in which voltage-sensitive fluorescent dyes are added to cardiomyocyte preparations before imaging at high spatio-temporal resolution \cite{Efimov2004}. It can be used to effectively obtain uncalibrated measurements of $V(\vec{x},t)$ in cardiac tissue across time. Although optical mapping can be challenging \textit{in vivo}, \textit{in vitro} experiments can provide very detailed insights into AP properties and cellular-level EP properties and gain insights into arrhythmic mechanisms \cite{Hansen2018}. In particular, optical mapping can be used to study the effect of anti-arrhythmic drugs on cardiomyocyte preparations \cite{Chowdhury2018}. These data were used to test EP-PINNs in an experimental setting.

\section{Material and Methods}
In this section, we provide details about the finite differences (FD) model used, in a variety of settings, to generate training and test data for EP-PINNs. We then introduce the EP-PINNs architecture, before giving details about each of the \textit{in silico} experiments in which EP-PINNs were deployed. We end this section by introducing the experimental data (\textit{in vitro} optical mapping) used for testing EP-PINNs. The code used for EP-PINNs implementation and the  generation of \textit{in silico} GT data is freely available from \href{https://github.com/martavarela/EP-PINNs}{github.com/martavarela/EP-PINNs}.

\subsection{Generation of \textbf{in silico} EP Data} \label{GT Generation}
We use an FD solver to generate \textit{in silico} cardiac EP data, $V_{GT}$, which we use to train and evaluate the performance of EP-PINNs. Simulations were carried out in two different geometries: in 1D, in a 2-cm 1D domain cable; and in 2D, in a square with a 1-cm side. We use in-house code written in Matlab 2020b (Mathworks, Natick, MA, USA) that relies on central FD and an explicit 4-stage Runge-Kutta method to solve the isotropic monodomain Aliev-Panfilov model, as defined in Eqs \ref{eq:AlievP_V} and \ref{eq:AlievP_W} and model parameters listed in Supplementary Table 1. All simulations use Neumann boundary conditions and set $dt = 5\e{-3}~TU$ and $dx = 100 \mu m$ as temporal and spatial steps. All simulations were run for $300~TU$, with the calculated $V_{GT}(\vec{x},t)$ field saved at every $1~TU$ and at every spatial step (every $100 \mu m$). We thus generate in total $1.4\e{4}$ and $7.0\e{5}$ data points for $V_{GT}(\vec{x},t)$, respectively in 1D and 2D. 

Figure \ref{fig:Fig1} shows example time frames of the generated $V$ maps in 2D. All APs were initialised by adding an external stimulus current $V_{stim} = 0.12~AU$  to the right-hand side of Eq \ref{eq:AlievP_V} for $1~TU$ in a sub-domain of the studied geometry (Figure \ref{fig:Fig1}). In 2D simulations, this includes both planar (Figure \ref{fig:Fig1}a) waves (emanating from a rectangular stimulus) and centrifugal waves (from a point-like excitation) (Figure \ref{fig:Fig1}b). Spiral waves were also created using the cross-field protocol: as an initial planar excitation propagates, a second planar excitation wave, orthogonal to the first, is initiated. When timed appropriately ($42~TU$ after the first stimulus, in our model), the second planar wave continuously curves as it moves to non-refractory tissue, giving rise to a sustained spiral wave (Figure \ref{fig:Fig1}c). 

The data generated by the FD model is treated as GT data and is used to both train and test EP-PINNs. As in past studies \cite{Lu2021DeepXDE:Equations, Karniadakis2021}, we train EP-PINNs on a case-by-case basis, with a small subset of the data for which parameter inference is going to be performed. This is in contrast to most supervised NNs, which are usually trained with large amounts of data acquired in varied circumstances.

Training data for EP-PINNs, $V_{GT} (\vec{x},t)$, are provided to the network as an input and used to minimise the data discrepancy loss function, $L_{data}$ in Equation \ref{eq:Loss}. They amount to 10-20\% of the total generated data, corresponding to a variable number of data points as detailed below and in Supplementary Table 2. The remaining data are withheld from EP-PINNs and used, in a post-processing step, to assess EP-PINNs' ability to reproduce cardiac APs, as detailed below.

\subsection{Architecture and Training of EP-PINNs}
EP-PINNs are designed, trained and deployed using the Python DeepXDE library \cite{Lu2021DeepXDE:Equations}. As in past successful implementations of PINNs \cite{Karniadakis2021, Lu2021DeepXDE:Equations}, we use a fully connected network architecture.

\begin{figure}[h]
\centering
\includegraphics[width=\columnwidth]{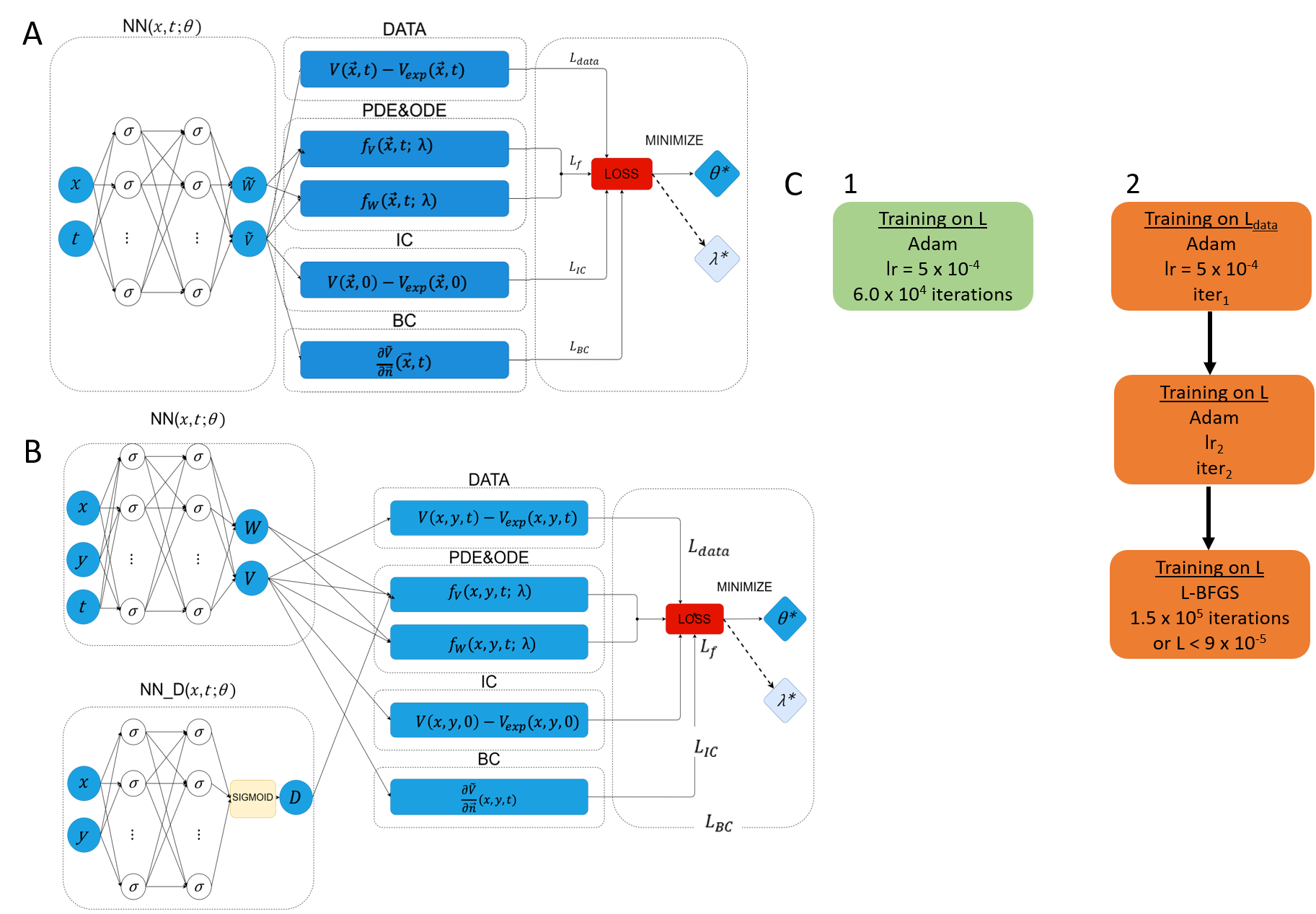}
\caption{Network architectures and training schemes for EP-PINNs used in this study. Neurons are represented by $\sigma$. In forward simulations only network parameters $\theta$ were estimated, whereas in the inverse setting one or more EP parameters ($\lambda$) were also calculated. \\
\textit{A} - Architecture used in all forward simulations and all inverse simulations (1D and 2D) with homogeneous EP parameters. \\
\textit{B} - Architecture used in all 2D forward simulations and all inverse simulations where $D$ was a spatially-varying field. \\
\textit{C} - Used training schemes. 1 was used in \textit{in silico} 1D problems, whereas training scheme 2 was used in 2D problems and for optical mapping experimental data. \\
The number of NN neurons and layers varied across different experiments, as did the learning rates (lr) and number of iterations (iter) in training scheme 2. Further details about the architecture and parameterisation of the NNs and training schemes can be found in Supplementary Table 2.}
\label{fig:Fig2}
\end{figure}

As detailed in Figure \ref{fig:Fig2}, EP-PINNs take as inputs the spatio-temporal points, $(\vec x, t)$, where they will estimate the main outputs: $V$ (and $W$). Experimental measurements of $V$, $V_{GT}(\vec x, t)$, are also provided to EP-PINNs as inputs in a (training) subset of $(\vec x, t)$. EP-PINNs minimise the hybrid physics-informed loss function described before (see Eqs \ref{eq:LossTerms} and \ref{eq:Loss}), by adjusting the network's weights and biases (collectively named $\theta$ in Figure \ref{fig:Fig2}). In inverse mode, EP-PINNs additionally adjust one or more parameters of the Aliev-Panfilov model (generically $\lambda$ in Figure \ref{fig:Fig2}). The number of layers and neurons used by EP-PINNs is adjusted to the domain size and the complexity of the problem at hand, as described below and detailed in Supplementary Table 2.

The optimisation approach used for EP-PINNs also depends on problem size. For 1D problems, we use Adam optimisation \cite{Kingma2015} - see Figure \ref{fig:Fig2}c. We empirically determined that, in 2D, EP-PINNs' performance improved when initially using Adam optimisation for only the data agreement term ($L_{data}$ in Eq \ref{eq:Loss}), followed by Adam optimisation for the full loss function and ending in a final phase of L-BFGS optimisation \cite{Liu1989OnOptimization} - see Figure \ref{fig:Fig2}c. The initial Adam training phases are used to rapidly approach the desired minimum and the final L-BFGS optimization phase helps the network converge faster towards it \cite{Lu2021DeepXDE:Equations}.

In the presence of spatially-varying EP parameters ($D$ in the current study), we use network architecture B (see Figure \ref{fig:Fig2}B). Here, $D(\vec{x},t)$ is estimated by a parallel NN, $NN_D$, with the same number of layers and neurons as the main NN. In this setup, $D$ is treated as a system variable (on par with $V$ and $W$) and its estimates directly contribute to the loss term that ensures the agreement with the EP equations: $L_{f_V}$ in Eq \ref{eq:Loss}. 

The hyperbolic tangent function ($tanh$) is used throughout as the differentiable activation function and Glorot initialisation from a uniform distribution is used for all weights \cite{Glorot2010}. Additionally, to minimize convergence problems caused by explosive gradients \cite{Lu2021DeepXDE:Equations} and enhance the NN's stability, we implemented an automatic reset of the training process when the losses at the first epoch of the training exceeded a predefined threshold. 

EP-PINNs were trained on a high performance machine with 1 RTX6000 GPU and 4 AMD EPYC 7742 CPUs. Typical training times varied between 15 min (for 1D problems) and 16h (for heterogeneous spiral wave problems), as detailed in Supplementary Table 2.
 
\subsubsection{Assessment of EP-PINNs Performance}

To assess the performance of EP-PINNs, we calculate, across all test points $N_{test}$, the root mean squared error ($RMSE$) for estimates of $V$:
\begin{equation}
    RMSE = \sqrt{\frac{1}{N_{test}} \sum_{i=1}^{N_{test}} (V(i) - V_{GT}(i))^2 } \label{eq:EqRMSE}
\end{equation}
The RMSE is, by construction, adimensional and in the same scale range as $V$. All experiments were repeated at least 5 times to probe the variability in RMSE. In  inverse mode, we additionally calculate the precision of the estimated model parameters using the standard deviation of the parameters estimated by EP-PINNs in these 5 different runs.

\subsection{Forward Solution of EP Models} \label{ForwardMethods}

\subsubsection{1D Cable Geometry}
We assessed the accuracy of EP-PINNs when solving the monodomain equation with the Aliev-Panfilov ionic model (forward problem) in the 20-cm cable during 70 TU (corresponding to 903 ms). We divided the GT data from the FD solver, $V_{GT}$, into test and training datasets using a 90-10\% split. This corresponds to 140 randomly chosen points across the temporal and spatial domains for training EP-PINNs and 1,260 points for testing. As for inverse 1D problems, EP-PINNs were implemented in this instance using architecture A and training scheme 1 (see Supplementary Table 2).

We used two different training setups:
\begin{enumerate}
    \item using only \textit{in silico} experimental $V_{GT}$ measurements as GT, as in Eq \ref{eq:Loss}. 
    \item using \textit{in silico} experimental points for $V_{GT}$ and $W_{GT}$ in the loss function, by adding an extra term: $L_{W_{GT}} = \frac{1}{N} \sum_{i=1}^{N} (W({x_i},t_i)- W_{GT_i}^2 $ to Eq \ref{eq:LossTerms}.
\end{enumerate}
Setup 1 more closely resembles an experimental setup, as the latent variable $W_{GT}$ is not usually measurable.

To gauge whether EP-PINNs performance depended on model parameter choice, we used EP-PINNs on GT data synthesised with two different sets of model parameters, as detailed in Supplementary Table 1.

We additionally evaluated the performance of EP-PINNs in \textit{in silico} data corrupted by noise. For this, we added to $V_{GT}$ zero-mean Gaussian noise with standard deviations of $0.05$, $0.10$, $0.50$ or $1.00~AU$ (with 1 AU being the approximate amplitude of an AP).

We also tested the performance of EP-PINNs in the presence of a reduced number of training data points. For this, we provided the network with $V_{GT}$ at \num{1e4}, \num{5e3}, \num{1e3} or \num{100} random training points within the 1D space-time domain (compared to 1.40\e{7} in usual conditions). We used architecture A with training scheme 1 in all 1D problems (see Supplementary Table 2).

\subsubsection{2D Rectangular Geometry} \label{2DForwardMethods}
We solved the forward problem in 2D, using the Aliev-Panfilov model in an isotropic and homogeneous 10-cm side square over 70 TU. We used $\num{1.4e5}$ randomly chosen data points (across the temporal and spatial domains) from the FD solver for the training of EP-PINNs (corresponding to 20\% of the total generated points) and reserved the remaining $\num{5.6e5}$ points to assess EP-PINNs' performance. Simulations were carried out in three scenarios: planar wave (Figure \ref{fig:Fig1}a), centrifugal wave, emanating from point-like excitation in a corner (Figure \ref{fig:Fig1}b) and a spiral wave (Figure \ref{fig:Fig1}c). Here and in the equivalent inverse setup, EP-PINNs were implemented using architecture A and training scheme 2 (see Supplementary Table 2).

The point-like excitation and spiral wave scenarios were also simulated in heterogeneous conditions, where a 2-mm side square within the spatial domain was assigned a permanent diffusion coefficient ($D_{lesion}=0.02~mm^2/TU$) lower than that of background tissue ($D_0=0.1~mm^2/TU$). Architecture B with training scheme 2 was used for all (forward and inverse) heterogeneous problems (see Supplementary Table 2). A sigmoid function was used in the NN dedicated to estimating $D(\vec{x},t)$ ($NN_D$ in Figure \ref{fig:Fig2}B) to account for the fact that $D(\vec{x},t)$ follows a binomial distribution: $D_0$ in healthy tissue and $D_{lesion}$ otherwise.

\subsection{Inverse Estimation of EP Parameters}
We withheld the value of one or more of the model parameters from EP-PINNs, which were instead estimated by it. These parameters were chosen for their comparatively simple biophysical interpretation, known susceptibility to both disease remodelling and pharmacological action, and the limited degree of mathematical coupling between them. They were: 
\begin{itemize}
    \item $a$, which is related to the tissue excitation threshold (see Eq \ref{eq:AlievP_V}); 
    \item $D$, the scalar diffusion coefficient (proportional to the electrical conductivity of the tissue, see Eq \ref{eq:AlievP_V});
    \item $b$, which controls APD, see Eq \ref{eq:AlievP_W}.
    \item $a$ and $D$ simultaneously;
    \item $b$ and $D$ simultaneously.
\end{itemize}

\subsubsection{Homogeneous 1D and 2D Geometries}
We solved the inverse problem in the same (1D or 2D) setup, architecture and training scheme as for the forward problems and using a similar division of randomly chosen data points for training and testing. We assumed that none of the selected parameters varied across time or space (except for $D$, in the heterogeneous problem described below) and used the values listed in Supplementary Table 1 as GT values. In 1D, we additionally investigated how EP parameter estimation was affected by experimental noise. For this, Gaussian noise ($\sigma = 0.05$ or $0.10~AU$) was added to the \textit{in silico} data as described for the forward mode in Section \ref{ForwardMethods}.

In addition to EP-PINNs robustness in the presence of experimental noise, we are also interested in its ability to cope with model uncertainty. Therefore, to assess EP-PINNs' ability to generalise beyond the model it is trained on, we additionally tested it on APs generated on a much more complex canine atrial EP model \cite{Varela2016}. These atrial APs are markedly different from those of the Aliev Panfilov model the EP-PINNs assumes, both in morphology and restitution properties. The canine atrial model data were synthesised using Matlab (\href{https://models.cellml.org/workspace/47c}{https://models.cellml.org/workspace/47c}) with central FD and explicit forward Euler schemes($dt = 5 \mu s$ and $dx = 100 \mu m$), with data saved at every spatial step and at every ms. Using this model, we created GT data for left atrial cells at 3 stages of AF-induced remodelling, which differed in APD. We tested, using the 1D model in inverse mode, whether EP-PINNs could identify the reduction in APD (detected as an increase in parameter $b$) in left atrial APs caused by increasing amounts of AF remodelling. 

\subsubsection{Estimation of EP Parameter Heterogeneities}
We assessed EP-PINNs' ability to recognize spatial heterogeneities in model parameters in 2D, as a test for EP-PINNs potential for identifying spatially-varying lesions such as fibrosis. For this, we used the same setup as in section \ref{2DForwardMethods}, with $D_0=0.1~mm^2/TU$ reduced to $D_{lesion}=0.02~mm^2/TU$ in a similar square region. As before, we estimated $D(\vec{x},t)$ on its own and simultaneously with either the $a$ or $b$ global model parameters.

\subsection{Parameter Estimation using Optical Mapping Data}
We tested EP-PINNs performance on \textit{in vitro} datasets using optical mapping data from neonatal ventricular rat myocyte preparations stained with a voltage-sensitive dye, as described in detail by Chowdhury \textit{et al} \cite{Chowdhury2018}.
Briefly, we used four time series (movies) of optical mapping images (field of view: $4.1 \cross 0.1~mm$, spatial resolution: $1.2~\mu m$, temporal resolution: 2 ms, duration: 300 ms). In two of these image series, ionic channel modulating drugs (E-4031 or nifedipine) had been administered at half maximal inhibitory concentration ($IC_{50}$): 772.2 nM for nifedipine and 243.4 nM for E-4031. The other two temporal image series consisted of matched control (baseline) preparations, to which no drug had been given.

We manually selected two square regions of interest (ROIs) with a side of $2.3~\mu m$ and with their centres $0.7~mm$ apart, in the same image location for each time series. We spatially averaged the optical signal over each ROI to obtain a signal trace across time. From this signal, we manually selected two consecutive APs and normalised the signal to the [0, 1] interval for consistency with the Aliev-Panfilov model. To improve the signal to noise ratio (SNR) of this trace, we applied a mean average filter twice, aligned and averaged the two APs over time to obtain a single higher-SNR AP. These pairs of post-processed APs were used as inputs to 1D EP-PINNs in inverse mode, with $b$ as the variable to be estimated. All model parameters were unchanged from those in Supplementary Table 1. We used EP-PINNs' architecture A with training scheme 2 to estimate $b$ 10 times for each setting. 116-200 points were used for training EP-PINNs and 29-50 to test it, as detailed in Supplementary Table 2.

We investigated in particular whether EP-PINNs could detect the effect on APD of E-4031 and nifedipine, which respectively block the hERG voltage-gated potassium channel ($I_{Kr}$ current) and the L-type calcium channel ($I_{CaL}$ current). Whereas E-4031 will extend APD (and thus shorten $b$ in the Aliev-Panfilov model - see Eqs \ref{eq:AlievP_V}, \ref{eq:AlievP_W}), nifedipine will have the opposite effect, decreasing APD (increasing $b$). We compared EP-PINNs' $b$ estimates for i) E-4031 vs. baseline and ii) nifedipine vs. baseline and assessed, using a t-test, whether the administered drugs significantly changed $b$ values.

\section{Results}
EP-PINNs successfully solved the monodomain equation coupled with the Aliev-Panfilov model in 1D and 2D, including in the presence of heterogeneities in D. In inverse mode, the proposed setup could also successfully perform EP parameter estimation from \textit{in silico} and \textit{in vitro} data, as described below.

\subsection{Forward Solution of EP Models}

\begin{figure}[h]
\centering
\includegraphics[width=\columnwidth]{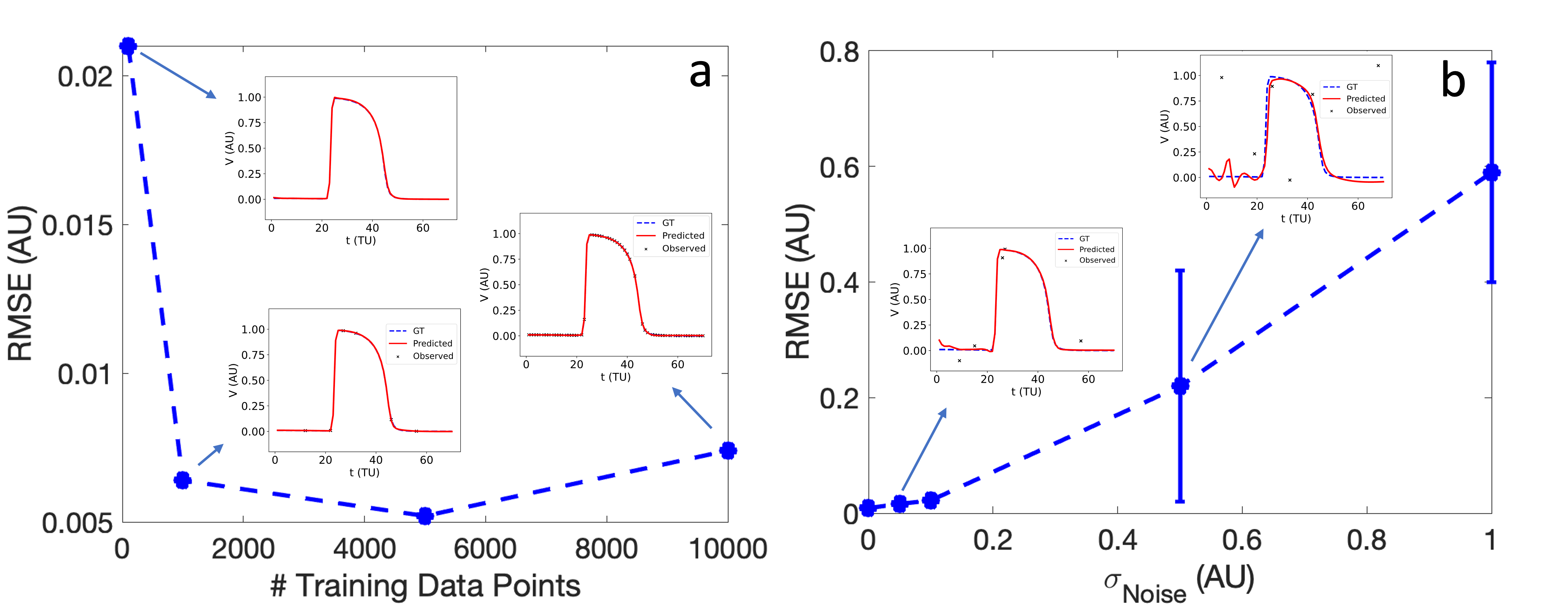}
\caption{Effect of the number of sampled experimental points (a) and experimental noise (b) on EP-PINNs $V$ estimates. The error in estimation is measured using the root mean square error (RMSE, see Eq \ref{eq:EqRMSE}) in forward mode in 1D. Representative $V(t)$ plots sampled at a random spatial location are shown as insets for some of the probed conditions.}
\label{fig:3}
\end{figure}

\subsubsection{1D Cable Geometry}
EP-PINNs accurately reproduced the features, morphology and conduction properties of the APs generated by the Aliev-Panfilov model (Figure \ref{fig:3}a)). We found that EP-PINNs could accurately simulate APs even in the absence of GT values for the latent variable $W$. Although the error was slightly increased in the absence of $W_{GT}$ ($RMSE = \num{6.0e-3} \pm \num{2.0e-3}$ vs $\num{9.0e-3} \pm \num{4.0e-3}$), the RMSE was still minimal (see top left inset in Figure \ref{fig:3}a) and the estimated $V(\vec{x},t)$ were visually indistinguishable from GT traces in both cases. As a consequence, in all other experiments described in this study, GT data for $W$ was not used to train EP-PINNs, whose training relied solely on $V_{GT}$.

We found that EP-PINNs could solve the model accurately in the presence of even small numbers of $V_{GT}$ points for training, with $RMSE \le \num{2.5e-2}$ even when trained with only 100 points (Figure \ref{fig:3}a). As expected, increasing noise in $V_{GT}$ led to increasing levels of error in $V$ estimates (Figure \ref{fig:3}b), but EP-PINNs were able to converge in the presence of Gaussian noise with a standard deviation below $0.5~AU$ (approx. 0.5 times the amplitude of an AP). When using a different set of parameters for the Aliev-Panfilov model (see Supplementary Table 1), EP-PINNs' accuracy was comparable, suggesting that the obtained EP-PINNs performance is robust to different biophysical model settings.

\subsubsection{2D Rectangular Geometry}
\begin{figure}[h]
\centering
\includegraphics[width=\columnwidth]{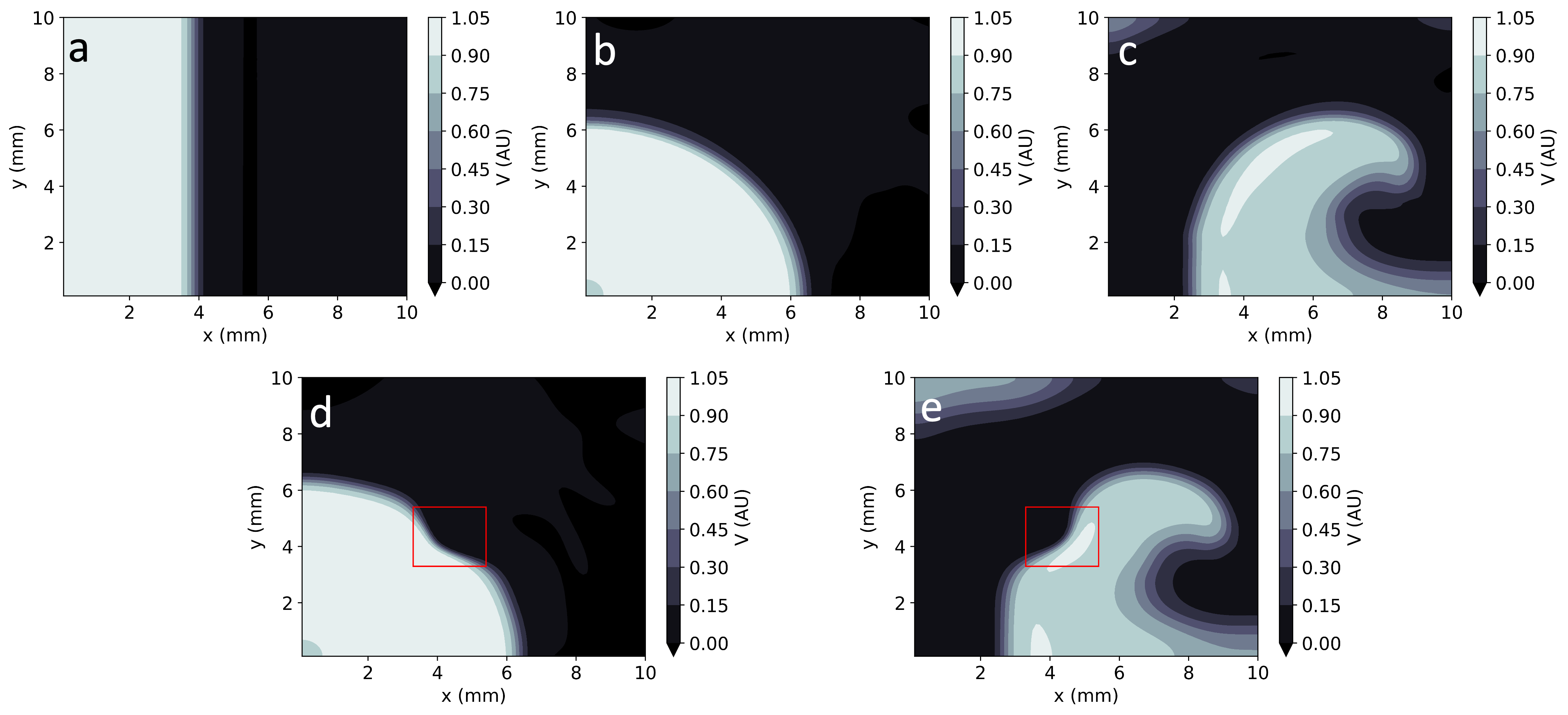}
\caption{EP-PINNs 2D forward solutions to the Aliev-Panfilov monodomain system for the same conditions as the GT data depicted in Figure \ref{fig:Fig1}. \textbf{a)} Planar wave; \textbf{b)} Centrifugal wave; \textbf{c)} Spiral wave; \textbf{d)} Centrifugal wave in the presence of a square heterogeneity in $D$; \textbf{e)} Spiral wave in the presence of a square heterogeneity in $D$.}
\label{fig:4}
\end{figure}

In homogeneous conditions, EP-PINNs were also able to reproduce AP propagation in 2D for planar, centrifugal and spiral waves (see Figure \ref{fig:4}a-c), with excellent accuracy ($RMSE < \num{3.0e-2}$ throughout). In the presence of heterogeneities in $D$, EP-PINNs were also able to accurately simulate APs with an RMSE of $\num{5.6e-3} \pm \num{6.6e-4}$ for centrifugal waves, which increased slightly to $\num{2.7e-2} \pm \num{4.3e-3}$ for spiral ones (Figure \ref{fig:4}a-c). In the spiral wave scenario, EP-PINNs found it most difficult to reproduce $V$ in the high wavefront curvature regions close to the spiral wave tip. 

Movies showing the propagation of APs in the 2D rectangular domain across time (for both GT and EP-PINNs solvers) can be seen in Supplementary Videos 1-5.

\subsection{Inverse Estimation of EP Parameters}

\subsubsection{Homogeneous 1D and 2D Geometries}
\begin{figure}[h]
\centering
\includegraphics[width=\columnwidth]{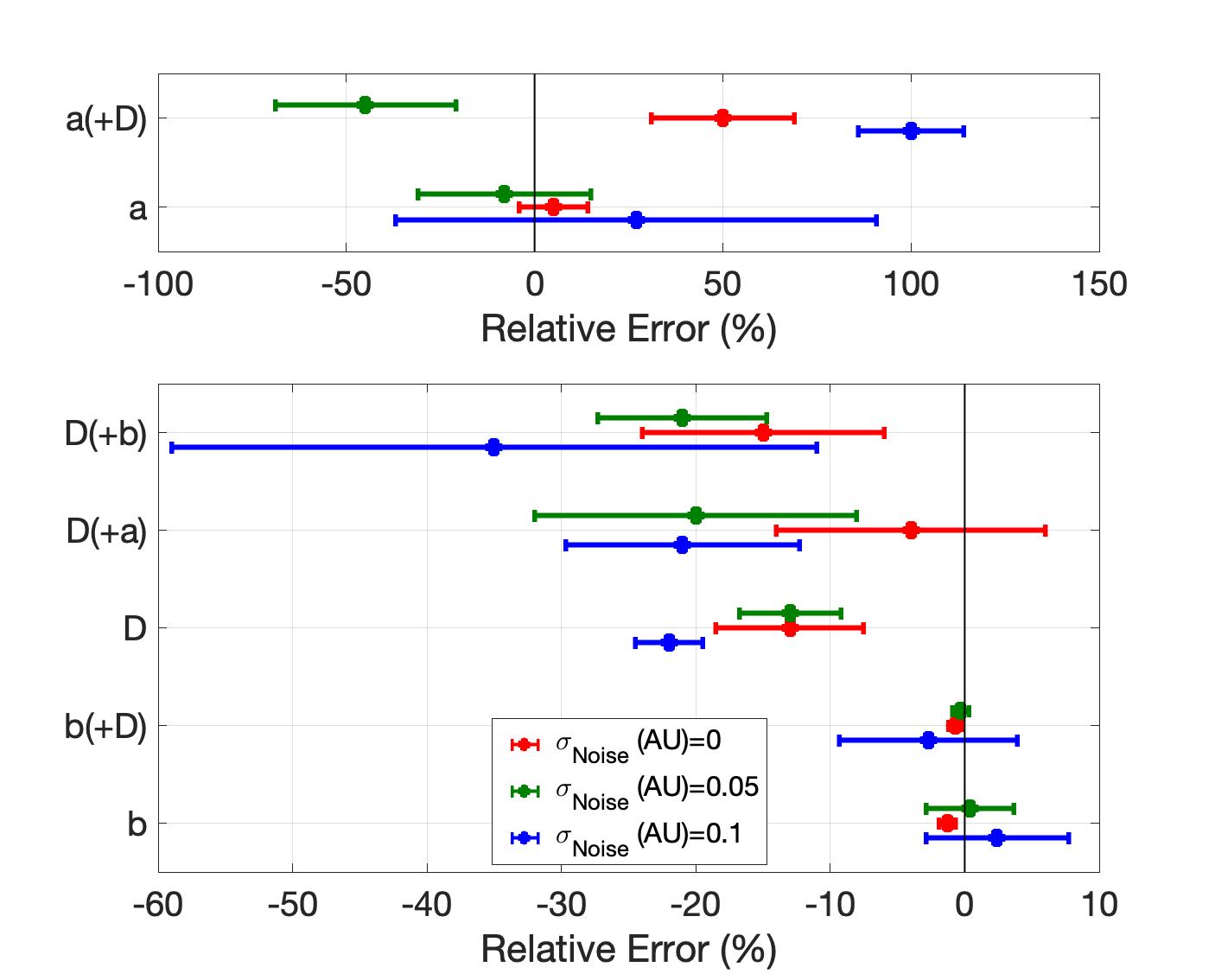}
\caption{Error in global EP parameter estimates by EP-PINNs in the inverse setting in 1D in the presence of different amounts of experimental noise. We show the relative error for $a$, $b$ and $D$, when estimated separately and in pairs.}
\label{fig:Fig5}
\end{figure}

EP-PINNs were able to estimate global model parameters in 1D the presence of varying degrees of noise, as detailed in Figure \ref{fig:Fig5}. As for the forward problem (Figure \ref{fig:3}), AP morphology and main properties were well reproduced even in the presence of large amounts of noise, with $RMSE < \num{9.0e-3}$ overall. Example plots of $V(t)$ in inverse mode in 1D are shown in Supplementary Figure 1.

When estimating only one model parameter, relative errors ($RE$) did not exceed on average $27\%$, even in the presence of noise. Estimates of $b$, which determines AP duration, were the most accurate ($|RE| < 3\%$), followed by $D$ ($|RE| < 35\%$), which EP-PINNs tended to underestimate. $a$ estimation was considerably more difficult. Joint estimation of two parameters led in general to less accurate parameter estimates (Figure \ref{fig:Fig5}), with an error as high as 100\% for $a$ when estimated jointly with $D$ in the presence of noise (see Figure \ref{fig:Fig5}). When performing simultaneous estimation of two parameters, no evidence of coupling between them was observed.

\begin{figure}[h]
\centering
\includegraphics[width=\columnwidth]{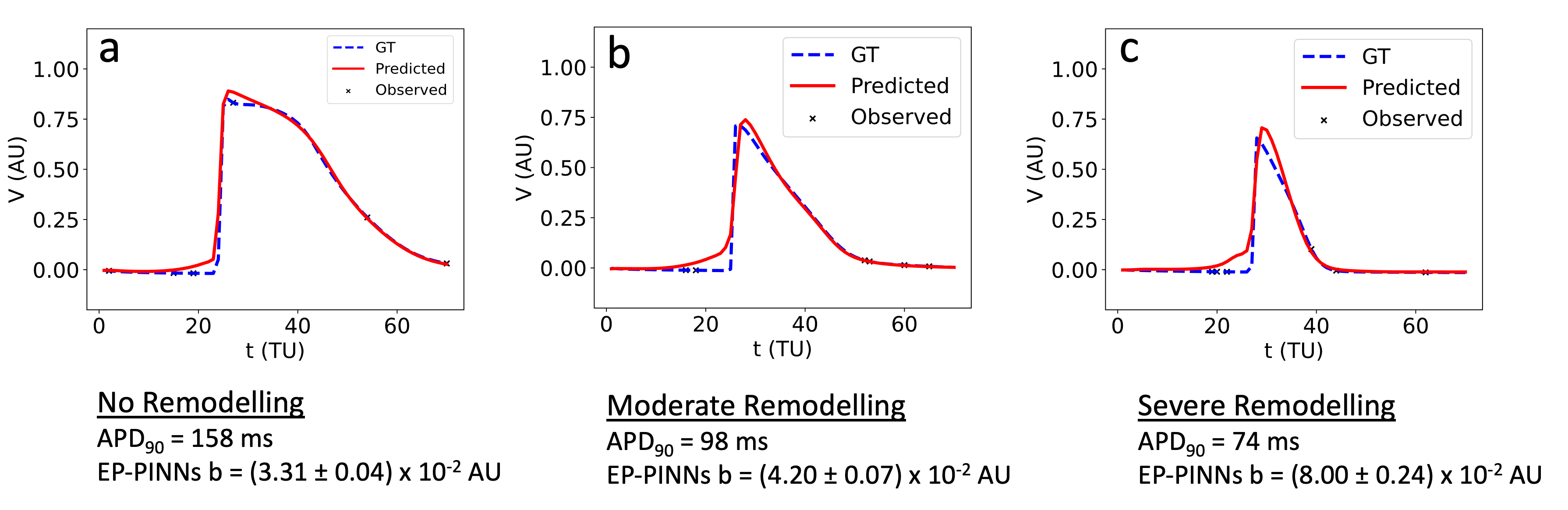}
\caption{1D inverse EP-PINNs solution for a detailed canine left atrial model in the conditions of: a) no remodelling; b) moderate AF remodelling; c) severe AF remodelling. Representative $V(t)$ plots are shown throughout, accompanied by the models $90\%$ APD and EP-PINNS estimates for $b$, a parameter inversely proportional to APD.}
\label{fig:Fig6}
\end{figure}

EP-PINNs were additionally able to perform robust parameter estimation on synthetic experimental data generated by a different EP model \cite{Varela2016}. When estimating $b$ in APs generated by a different atrial EP model, it correctly inferred that APD is reduced (i.e. $b$ is increased) for increasing degrees of AF remodelling, as shown in Figure \ref{fig:Fig6}. Moreover, the main AP features were well reproduced, with small discrepancies related to the differences between the two models (Figure \ref{fig:Fig6}). Interestingly, the solution proposed by EP-PINNs consistently shows a less steep depolarization than expected from either Aliev-Panfilov model and the detailed canine model. This mismatch is likely to be a consequence of the EP-PINNs' adjustment to slightly different AP morphologies from those in the Aliev-Panfilov model in its loss function. This suggests that the cross-model estimation of parameters related to excitability (e.g. $a$) may not be as successful as parameters related to APD (such as $b$).

\begin{figure}[h]
\centering
\includegraphics[width=\columnwidth]{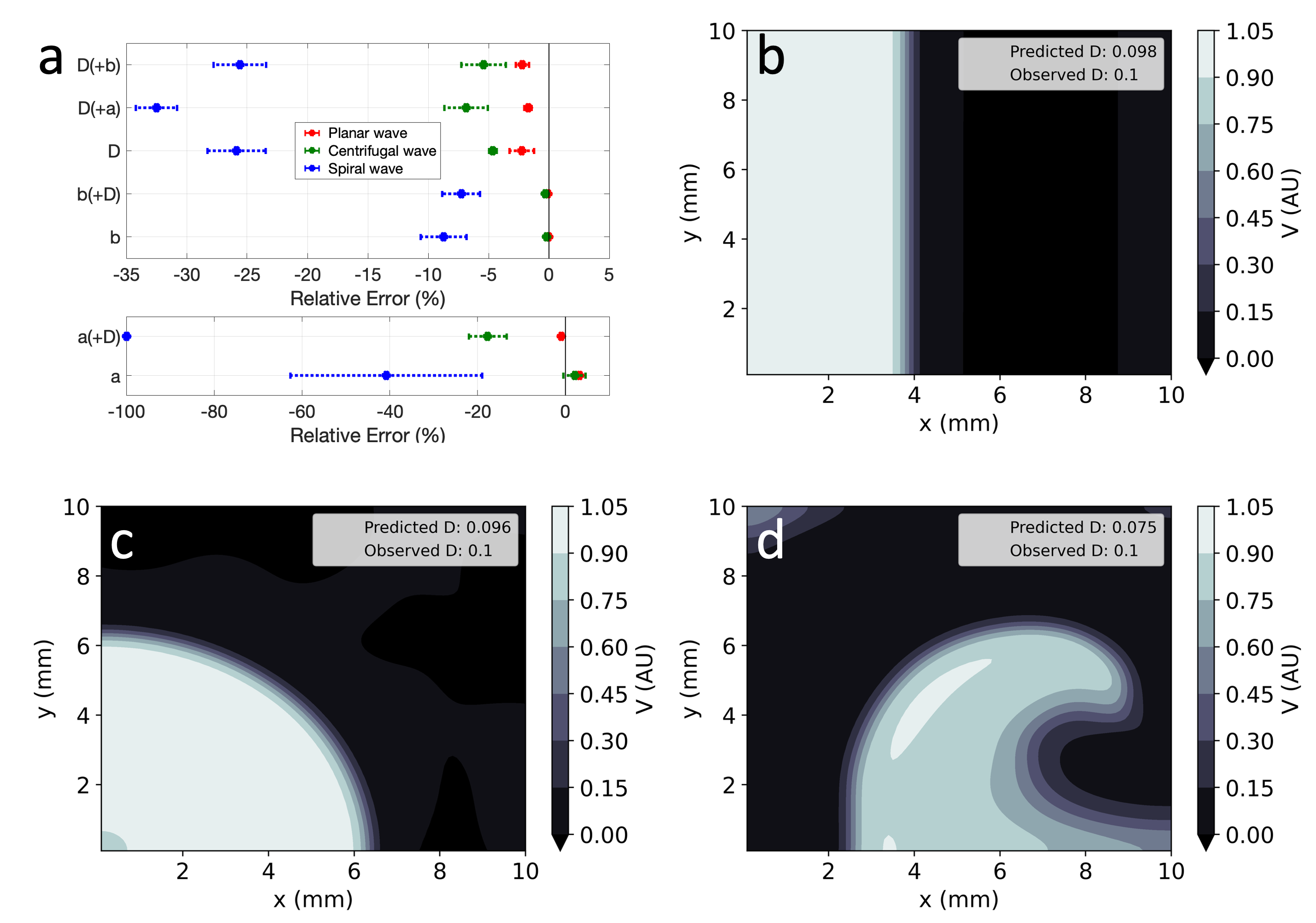}
\caption{EP-PINNs inverse solution in homogeneous conditions in 2D. a) Relative error for global estimates of: $a$ and $D$ and $b$ and $D$, when estimated separately or simultaneously. b-d) Corresponding representative $V$ maps for: planar wave (b), centrifugal wave (c) and spiral wave (d). Compare panels b-d to the corresponding GT in Figure \ref{fig:Fig1}a-c and the forward solutions in Figure \ref{fig:4}a-c.}
\label{fig:Fig7}
\end{figure}

Global parameter estimation in 2D was again successful, for both unidirectional propagation and spiral wave conditions, as demonstrated in Figure \ref{fig:Fig7}. As in 1D, $V$ was also correctly reproduced in the different analysed conditions (Figure \ref{fig:Fig7}b-d), with $RMSE < 2.3\e{-2}$ throughout. Across all experiments, parameter estimation and $V$ reconstruction were most successful for planar wave conditions followed by centrifugal waves and less accurate for spiral waves, as shown in Figure \ref{fig:Fig7}a). The comparatively worse performance of EP-PINNs in spiral wave conditions may be caused by the spatial dependency of wave front curvature in this setting.

Errors were largest when estimating two parameters simultaneously, with EP-PINNs again struggling to estimate $a$, especially when in conjunction with $D$ ($|RE|_a <100\%$). Estimates of $b$ were once again the most accurate ($|RE|_b<8\%$) and, as in 1D, EP-PINNs tended to underestimate $D$ across all settings and to underestimate all parameters in the spiral wave scenario.

\subsubsection{Estimation of EP Parameter Heterogeneities}

\begin{figure}[h]
\centering
\includegraphics[width=\columnwidth]{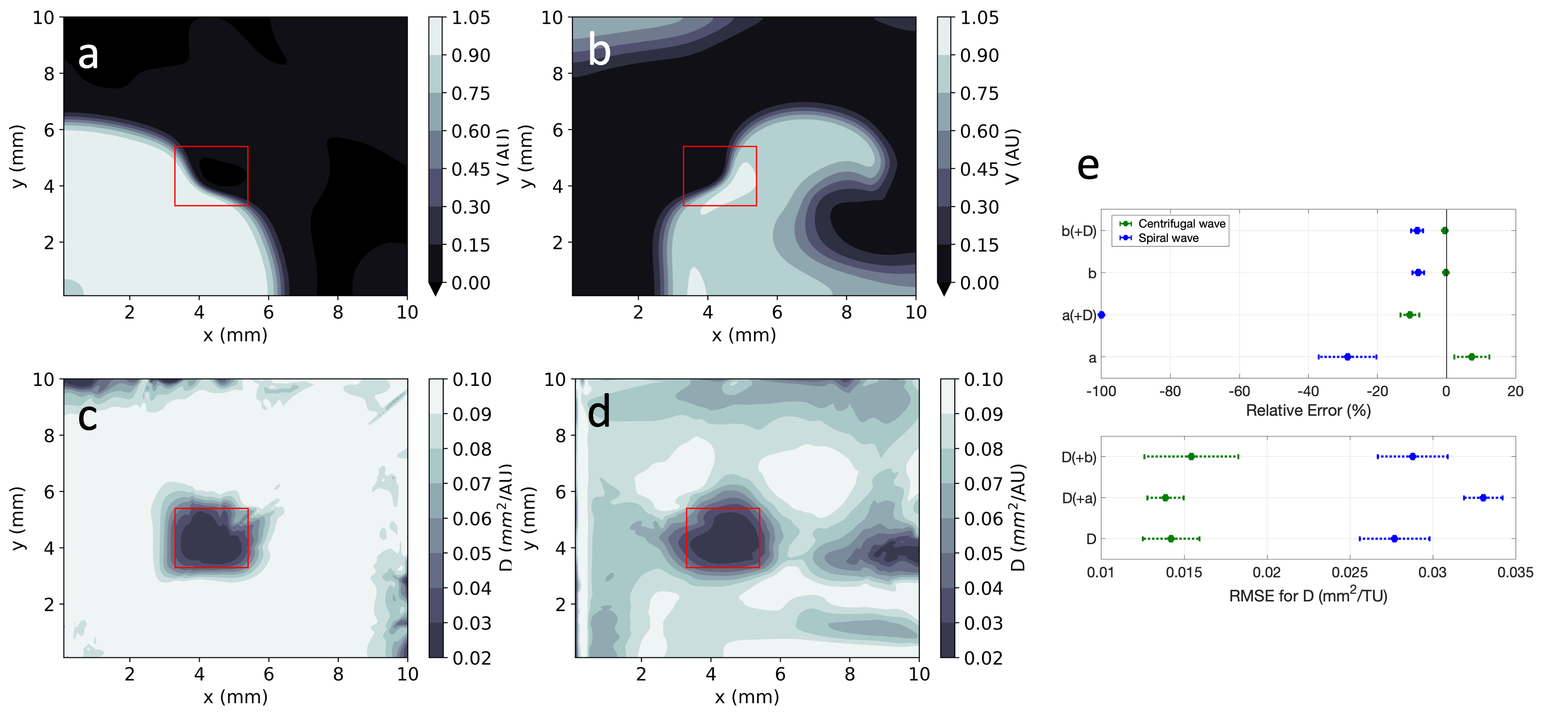}
\caption{EP-PINNs 2D inverse solution in the presence of heterogeneities in $D$. Maps showing representative $V$ and $D$ estimates for: a, c) centrifugal wave and b, d) spiral wave. e) Error for global estimates of $a$ and $b$ and RMSE for estimates of $D$ across the 2D domain, for all estimated parameter combinations. 
Compare panels a and b to the corresponding GT in Figure \ref{fig:Fig1}d-e and the forward solutions in Figure \ref{fig:4}d-e.}
\label{fig:Fig8}
\end{figure}

EP-PINNs were able to estimate $D$ on a pixel-by-pixel basis with remarkable accuracy, as demonstrated in Figure \ref{fig:Fig8}c,d, accurately identifying the low $D$ region. $RMSE_D$ was consistently below $3.5\e{-2}$ (Figure \ref{fig:Fig8}e) and was lower for the centrifugal wave case than the spiral wave one.
As before, $V$ was similarly well reproduced (Figure \ref{fig:Fig8}a,b), especially in the centrifugal wave scenario (see Figure \ref{fig:Fig8}b,d), with  $RMSE<3.0\e{-2}$. In the spiral wave scenario, EP-PINNs found the estimation of $D$ hardest near the spiral tip (see Figure \ref{fig:Fig8}d), where the high wavefront curvature may resemble the wavefront bending caused by low $D$ regions.

Using architecture B, global estimation of $a$ and $b$ was also possible in the presence of the heterogeneous $D$ field, keeping the same trends as in the 1D and 2D homogeneous cases (Figure \ref{fig:Fig8}e). As before, no evidence of coupling between the simultaneously estimated parameters was found. In detail:
\begin{itemize}
    \item $b$ was very accurately estimated ($|RE|<10\%$), whereas $a$ estimates had a larger error ($|RE|<100\%$);
    \item estimating $a$ and $b$ simultaneously with the $D$ field led to small decreases in accuracy overall;
    \item estimating $V$ and $D$ in the spiral wave setting was more difficult than in the presence of a centrifugal wave.
\end{itemize}

\subsection{Parameter Estimation using Optical Mapping Data}
\begin{figure}[h]
\centering
\includegraphics[width=\columnwidth]{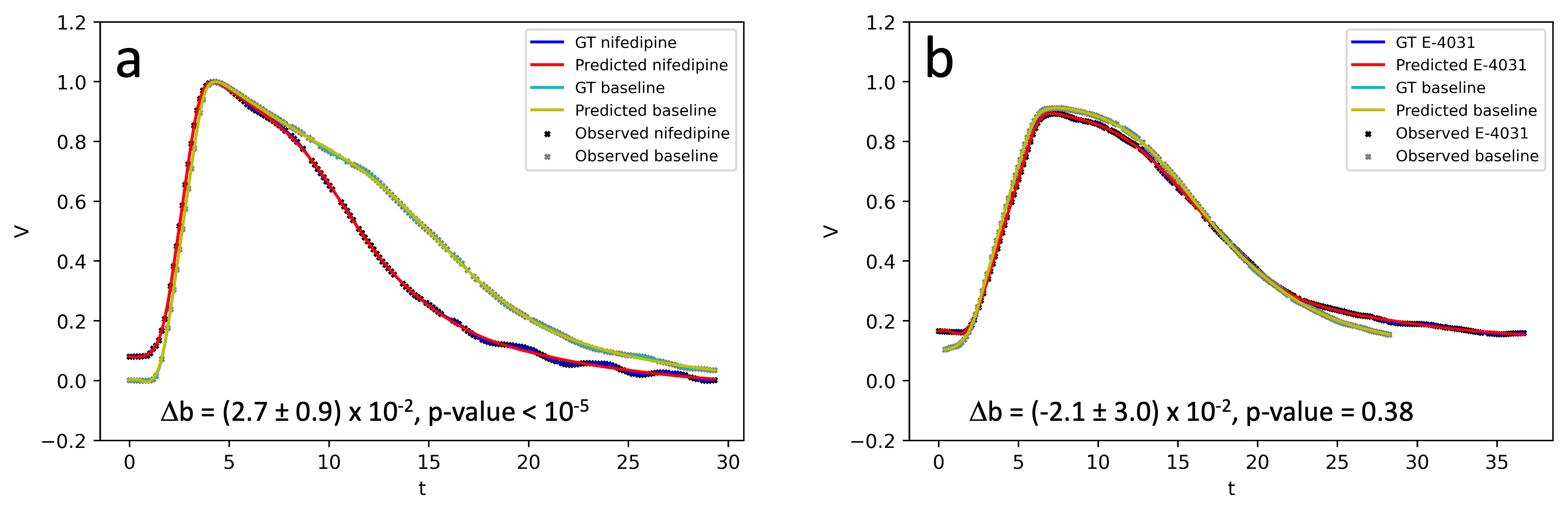}
\caption{1D inverse EP-PINNs solution for experimental optical mapping data, in the presence of a) $I_{CaL}$ channel blocker nifedipine and b) $I_{Kr}$ channel blocker E-4031. ($\Delta b$ refers to the change in model parameter $b$ in the presence of the drug when compared to the baseline value. Each AP was acquired in separate datasets and juxtaposed in the figure to allow visual comparisons.}
\label{fig:Fig9}
\end{figure}

Using optical mapping signals, EP-PINNs were able to accurately reproduce experimental APs and identify the actions of nifedipine and E-4031, correctly estimating that they respectively reduce and increase APD (Figure \ref{fig:Fig9}). The reduction in APD caused by nifedipine, an $I_{CaL}$ blocker, was detected by EP-PINNs as a significant increase in $b$ in the Aliev-Panfilov model ($\Delta b = (2.7 \pm 0.9)\e{-2}$, $p<10^{-5}$). The E-4031-induced increase in APD was more subtle ($\Delta b = (-2.1 \pm 3.0)\e{-2}$) and non-significant ($p=0.38$). The reduced effect of E-4031 in these data is consistent with the modest role $I_{Kr}$, the current blocked by E-4031, is expected to play in rodent APs \cite{Odening2021ESCResearch}.

\section{Discussion}
We present EP-PINNs, a successful framework to estimate EP parameters from measurements of trans-membrane potential $V$. We demonstrate EP-PINNs ability to accurately reproduce AP propagation in 1D and 2D in the presence of very sparse experimental measurements, experimental noise and model uncertainty. EP-PINNs can also estimate, for 1D and 2D \textit{in silico} and \textit{in vitro} data, global markers of APD, excitation threshold and/or conductivity (diffusion coefficient, $D$). We additionally show that EP-PINNs are further capable of identifying heterogeneities in EP parameters, such as $D$, even in arrhythmic conditions, showcasing their potential for clinically useful applications. 

\subsection{Forward Solution of EP Models}

EP-PINNs offer a flexible and easy to implement framework for parameter estimation in EP. Sahli-Costabal \textit{et al} and Grandits \textit{et al} had already demonstrated PINNs' potential in cardiac EP by estimating high-resolution left atrial AT and CV maps in sinus rhythm conditions using a simple activation-only biophysical model \cite{SahliCostabal2020, Grandits2021}. We extend PINNs' applications in EP by applying them to a more complex biophysical model, the monodomain Aliev-Panfilov model \cite{Aliev1996}, which also captures restitution properties through the inclusion of a latent (non-measurable) variable, $W$. For the first time, we use a PINNs framework for the simulation of arrhythmic conditions, such as spiral waves, and for the estimation of parameters unrelated to the AT of $V$. 

Importantly, we show EP-PINNs' are able to reproduce APs and perform parameter estimation in the absence of any data for $W$, which is not available experimentally. This bodes well for the deployment of PINNs for even more complex EP models, which use a higher number of latent variables to model individual ionic channels. This possibility is also supported by the work of Yazdani \textit{et al}, who successfully used PINNs for parameter estimation across several biological systems described by large sets of coupled ODEs \cite{Yazdani2020}.

We demonstrate PINNs' ability to describe AP dynamics in several circumstances. In 1D, EP-PINNs were able to reproduce APs even in the presence of very reduced amounts of experimental data (Figure \ref{fig:3}a) and large amounts of noise (Figure \ref{fig:3}b). PINNs' incorporation of explicit biophysical equations in the NN's loss function acts as an effective regulariser in EP problems, as demonstrated before in many other physical systems \cite{Raissi2019, Karniadakis2021}. We note that these inherent regularisation properties allow PINNs to be trained with much lower amounts of training data than conventional NNs. The main drawback is the need for a comparatively time-intensive training on a case-by-case basis, compared to the global training usually employed with supervised NNs.

In 2D, we were able to accurately replicate AP dynamics for planar, centrifugal and planar waves, even in the presence of heterogeneities in the diffusion coefficient (see Figures \ref{fig:4} and \ref{fig:Fig8}). We found that a more sophisticated training scheme and, for spiral waves, an increased NN capacity (5 layers of 64 neurons vs. 4 layers of 32 neurons, see Supplementary Table 2) were necessary for convergence in these large and complex problems. This more complex setup could, of course, have been used to solve the simpler 1D problems, through a trade-off between computational time and the convenience of a one-size-fits-all EP-PINNs approach.

\subsection{Inverse Estimation of EP Parameters}
It is in inverse mode, when estimating model properties from sparse measurements of $V$, that the EP-PINNs framework showcases its usefulness. Parameter estimation is an important topic in EP, as it is essential for both the personalisation of models and for the understanding of the effect of pathology and drugs on APs. Although parameter estimation in EP has been extensively discussed as a means of reducing the uncertainty associated with current biophysical models \cite{Clayton2020}, NNs had not yet been used for dedicated model parameter estimation in EP. This contrasts with the more common use of NNs as efficient solvers of EP systems (in a similar fashion to the EP-PINNs forward mode in the current study) \cite{Cantwell2019,Kashtanova2021,Fresca2020DeepElectrophysiology}.

Using EP-PINNs, we were able to estimate, with different degrees of accuracy, three different biophysical parameters, each controlling, in an almost uncoupled manner, different observable properties of the system: APD (through $b$), excitability (through $a$) and conduction velocity (through $D$). In both 1D and 2D (homogeneous and heterogeneous) problems, the network was highly successful at estimating $b$, but struggled with $a$, especially when estimating it in tandem with $D$. This is likely to reflect a dependency between EP-PINNs' inverse mode effectiveness and the solution type that is probed experimentally. Indeed, when compared to $b$, the experimental inputs to the network ($V_{GT}(\vec{x},t)$), depend little on $a$ providing the initial stimulus is supra-threshold. An exception could have been the spiral wave scenario, whose properties (e.g. the distance between spiral arms) depend strongly on model parameters such as $a$. Model parameters are more strongly coupled in the properties of spiral solutions, however, explaining the consistently lower accuracy of EP-PINNs' estimates in this scenario (Figure \ref{fig:Fig7}d and Figure \ref{fig:Fig7}a). Parameter estimation in the spiral wave scenario may be improved when EP-PINNs are trained in longer time series, in which the spiral wave tip samples more of the spatial domain.

These issues underlie the difficulties of finding a single experimental design that allows for simultaneous accurate estimation of several EP parameters. A solution for this problem may be the training of PINNs using data from the same system acquired in different experimental conditions, perhaps by training separate NNs in parallel with a combined loss function, in an analogous manner to architecture B in this study (see Figure \ref{fig:Fig2}).

EP-PINNs were additionally able to identify heterogeneities in $D$ across the 2D domain (see Figure \ref{fig:Fig8}), by estimating $D$ in a separate parallel NN which shared terms of the loss function ($L_{data}$, see Eq \ref{eq:LossTerms}) with the main NN. Characterising heterogeneities in EP parameters from electrical measurements is an interesting problem from a clinical point of view, as arrhythmias such as AF are often accompanied by heterogeneous changes in EP properties. Of these, regions of dense fibrosis (often modelled as areas with a reduced $D$ \cite{Roy2018}) are a promising candidate for personalised ablation sites \cite{Roy2020IdentifyingAtrium}, which may increase the overall efficacy of these procedures. EP-PINNs are thus well placed to help locate these putative ablation sites by identifying spatial heterogeneities in EP parameters such as $D$. 

The inputs to the current EP-PINNs implementation, $V(\vec{x},t)$ are not, however, measurable clinically. Future work will modify EP-PINNs to instead perform parameter inference from extracellular electrical potentials, $\phi_e$, which are regularly measured during clinical procedures using contact electrodes. $\phi_e$ can be interpreted as a weighted spatial convolution of the $\vec \nabla . (D \vec \nabla V)$ term in Eq \ref{eq:AlievP_W} \cite{Varela2014,Plonsey2007}, making the identification of localised EP changes more difficult. To take this into account, EP-PINNs designed for $\phi_e$ analysis may benefit from a move away from the current fully-connected architecture to incorporate, for example, convolutional layers. 

\subsection{Parameter Estimation using Optical Mapping Data}
An important point for future clinical applications of EP-PINNs is its ability to generalise beyond the details of the setup it is trained on. We showed that the proposed EP-PINNs implementation is model-agnostic, as it was able to perform robust parameter inference on \textit{in silico} data generated by a much more complex atrial EP model \cite{Varela2016} than the 2-variable ventricular one used in its loss function. In particular, EP-PINNs were able to correctly identify the decrease in APD (manifest as an increase in $b$) that is associated with increasing degrees of AF-induced remodelling in this model (see Figure \ref{fig:Fig6}).

In contrast to most previous PINN studies \cite{Raissi2019,Yazdani2020,Lu2021DeepXDE:Equations,SahliCostabal2020,Arthurs2021ActiveEquations}, we complemented the \textit{in silico} studies with an assessment of the EP-PINNs performance on experimental biological data. Despite requiring a proportionally higher amount of training data than \textit{in silico} experiments, EP-PINNs were able to cope well with the noise and artefacts unavoidably present in experimental data to identify the effect on APD of two different drugs: an $I_{Kr}$ blocker and an $I_{CaL}$ blocker. As for the data generated by a different mathematical model (Figure \ref{fig:Fig6}), EP-PINNs coped well with differences between the experimental data and the Aliev-Panfilov model, namely in resting membrane potential (see Figure \ref{fig:Fig9}). This ability to generalise well to data with different characteristics could be due to the use of PDE as a soft constraint (a term in the loss function) in the EP-PINNs framework, as well as the lack of assumptions about the distributions from which data come from.

As demonstrated for the \textit{in silico} tests, the EP-PINNs framework can easily be extended to simultaneously infer the effect of drugs on more than one EP parameter, which may be useful for the characterisation and safety assessments of anti-arrhythmic drugs. These applications may further benefit from the training of EP-PINNs on more complex biophysical models, to obtain a more fine-grained characterisation of potential pharmacological (or pathological) effects.

\subsection{Limitations and Future Plans}

The current study aims to demonstrate the potential of PINNs within EP, as an initial necessary step towards clinical applications of this method. As such, we only trained EP-PINNs using one comparatively simple EP model, in 1D and 2D scenarios and for the estimation of a small number of EP parameters. We additionally did not test EP-PINNs in the chaotic or pseudo-chaotic scenarios of spiral wave break-up \cite{Fenton1998}, which may be relevant for some arrhythmias.

Generalisations of the proposed framework to 2D/3D geometries representative of cardiac chambers, anisotropic conditions and more detailed EP biophysical models can be achieved by further increasing the capacity of the deployed EP-PINNs, with a concurrent increase in computational resources. However, promising and less resource-intensive applications for EP-PINNs may be the characterisation of pharmacological effects on AP or the identification of heterogeneities in EP properties from EGM signals.

\section*{Conflict of Interest Statement}
The authors declare that the research was conducted in the absence of any commercial or financial relationships that could be construed as a potential conflict of interest.

\section*{Author Contributions}
CHM and AO wrote the EP-PINNs software, conducted the experiments and analysed the results. RAC acquired the experimental data and oversaw its analysis. AAB, EU, RAC and NSP contributed to the study design and the revision of the results. MV designed the study, wrote the GT data generation software, critically analysed the results and drafted the manuscript. All authors critically reviewed the manuscript and gave final approval for publication.

\section*{Funding}
This work was supported by the  British Heart Foundation (RE/18/4/34215, RG/16/3/32175, PG/16/17/32069 and Centre of Research Excellence), the Imperial-TUM Seed Funding, the National Institute for Health Research (UK) Biomedical Research Centre and the Rosetrees Trust through the interdisciplinary award "Atrial Fibrillation: A Major Clinical Challenge".

\section*{Acknowledgments}
We thank the Imperial College Research Computing Service (DOI: \href{https://doi.org/10.14469/hpc/2232}{10.14469/hpc/2232}) for help with the computational setup of the project and technical support.

\section*{Data Availability Statement}
All data supporting the findings of this study are available within the article and from the corresponding author on reasonable request.

\bibliographystyle{styl} % for Health, Physics and Mathematics articles
\bibliography{refs}

\end{document}